\documentclass[aps,prb,a4paper,superscriptaddress,twocolumn,showpacs,amsmath,amssymb]{revtex4} 
\usepackage{graphicx,epsfig}
\usepackage[usenames]{color}

\allowdisplaybreaks
\begin{document}

\newcommand{\s}{\sigma}
\newcommand{\up}{\uparrow}
\newcommand{\dw}{\downarrow}
\newcommand{\h}{\mathcal{H}}
\newcommand{\g}{\mathcal{G}^{-1}_0}
\newcommand{\D}{\mathcal{D}}
\newcommand{\A}{\mathcal{A}}
\newcommand{\K}{\textbf{k}}
\newcommand{\Q}{\textbf{q}}
\newcommand{\T}{\tau_{\ast}}
\newcommand{\io}{i\omega_n}
\newcommand{\eps}{\varepsilon}
\newcommand{\+}{\dag}
\newcommand{\su}{\uparrow}
\newcommand{\giu}{\downarrow}
\newcommand{\0}[1]{\textbf{#1}}
\newcommand{\ca}{c^{\phantom{\dagger}}}
\newcommand{\cc}{c^\dagger}
\newcommand{\da}{d^{\phantom{\dagger}}}
\newcommand{\dc}{d^\dagger}
\newcommand{\be}{\begin{equation}}
\newcommand{\ee}{\end{equation}}
\newcommand{\bea}{\begin{eqnarray}}
\newcommand{\eea}{\end{eqnarray}}
\newcommand{\ba}{\begin{eqnarray*}}
\newcommand{\ea}{\end{eqnarray*}}
\newcommand{\dagga}{{\phantom{\dagger}}}
\newcommand{\bR}{\mathbf{R}}
\newcommand{\bQ}{\mathbf{Q}}
\newcommand{\bq}{\mathbf{q}}
\newcommand{\bqp}{\mathbf{q'}}
\newcommand{\bk}{\mathbf{k}}
\newcommand{\bh}{\mathbf{h}}
\newcommand{\bkp}{\mathbf{k'}}
\newcommand{\bp}{\mathbf{p}}
\newcommand{\bRp}{\mathbf{R'}}
\newcommand{\bx}{\mathbf{x}}
\newcommand{\by}{\mathbf{y}}
\newcommand{\bz}{\mathbf{z}}
\newcommand{\br}{\mathbf{r}}
\newcommand{\Ima}{{\Im m}}
\newcommand{\Rea}{{\Re e}}
\newcommand{\Pj}[2]{|#1\rangle\langle #2|}
\newcommand{\ket}[1]{|#1\rangle}
\newcommand{\bra}[1]{\langle#1|}
\newcommand{\fract}[2]{\frac{\displaystyle #1}{\displaystyle #2}}
\newcommand{\Av}[2]{\langle #1|\,#2\,|#1\rangle}
\newcommand{\eqn}[1]{(\ref{#1})}
\newcommand{\Tr}{\mathrm{Tr}}
\title{Variational Approach to transport in quantum dots}
\author{Nicola Lanat\`a}
\affiliation{} 
\affiliation{University of Gothenburg, SE-412 96 Gothenburg, Sweden}
\date{\today} 
\begin{abstract}
We have derived a variational principle that defines the nonequilibrium  steady state transport across a correlated impurity (mimicking e.g. a quantum dot) coupled to biased leads.  
This variational principle has been specialized to a Gutzwiller's  variational space, and applied to the study of the simple single-orbital Anderson impurity model 
at half-filling, finding a good qualitative accord with the observed behavior in quantum dots for the expected regime of values of the bias.
Beyond the purely theoretical interest in the formal definition of a variational principle in a nonequilibrium problem, the particular methods proposed have the important advantage to be simple and flexible enough to deal with more complicated systems and variational spaces.
\end{abstract}

\maketitle

\section{Introduction}

Nanocontacts of quantum dots, single molecules or atoms, and nanowires are ideal candidates to realize electronic devices where a 
source-drain current across the contact can be magnetically controlled.   
Indeed, because of the low dimensionality of the contact region, electronic correlations  
grow in strength and may stabilize a local magnetism that influences electron tunneling. The Kondo-like zero-bias anomalies 
first observed in quantum dots~\cite{GoldhaberGordon} are just the simplest manifestation of such a local magnetism, but one 
can foresee even more spectacular phenomena, like giant magnetoconductance.~\cite{Smogunov15}

From the theory side, this is a complicated problem first of all because electronic correlation is the main actor and is 
difficult to treat, and secondly because the inelastic tunneling spectrum requires full out-of-equilibrium calculations. 
Many complementary techniques have been used to characterize the nanocontact at equilibrium. For instance ab initio 
Local Density Approximation (LDA) calculations can provide the electronic structure and predict whether magnetism could 
indeed be stabilized,~\cite{Smogunov15,Smogunov10,Smogunov12}
at least at the mean field level. Inclusion of quantum fluctuations requires many-body techniques, like numerical renormalization 
group,~\cite{Wilson,Bulla-Review-of-Modern-Physics} which are often applied to oversimplified models, like the 
single-orbital Anderson impurity model, although there are recent attempts to join together the two 
approaches.~\cite{Costi-PRL,Jacob-PRL,Smogunov-Nature-Materials} Unfortunately, out of equilibrium properties are much more 
difficult to study. Apart from many-body Keldish perturbation theory,~\cite{Schoeller} many sophisticated numerical techniques have been 
developed in recent years to cope simultaneously with out-of-equilibrium and strong correlations.~\cite{Han,MuhlbacherandRabani,FBAnders,WeissEckelThorwartEgger,WernerMillis,schiro} However, given the complexity of the  
electronic structure that may arise at a nanocontact e.g. of a molecule or a bridging transition metal atom, 
it would be desirable to have at disposal approximate techniques simple and flexible enough  to deal with realistic situations 
otherwise prohibitive with more accurate numerical approaches, as those previously mentioned. 

In this paper we shall propose an out-of-equilibrium extension of the conventional variational approach, and, on such basis, an out-of-equilibrium extension of
the Gutzwiller approximation~\cite{Gutzwiller1,Gutzwiller2} for correlated 
electron systems.

The paper is organized as follows. In Sec.~\ref{theproblem} we briefly 
introduce the Hershfield formulation of the nonequilibrium steady state problem 
in quantum dots. In Sec.~\ref{theresonantmodeloutofequilibrium} we introduce the concept
of scattering operators and derive some results related with the 
resonant model. In Sec.~\ref{thebiasasanorderparameter} 
and~\ref{mainsec} we derive the variational 
principle that defines the nonequilibrium steady state transport across a finite 
junction (e.g. a quantum dot) coupled to biased infinite leads. 
In Sec.~\ref{theconceptofquasiparticlesoutofequilibrium} we 
formulate a Fermi-liquid assumption for the system in the low energy/temperature/bias
regime. 
In Sec.~\ref{geqsec} we introduce very briefly the standard Gutzwiller variational
method for the single band Anderson impurity model in equilibrium.
In Sec.~\ref{thegutzwillerapproximationoutofequilibrium} we propose 
a generalization of the Gutzwiller variational method to nonequilibrium.
Finally, Sec.~\ref{conclusions} is devoted to the conclusions.

\section{The problem}\label{theproblem}

We consider two biased macroscopic leads described by non-interacting 
electrons coupled 
to a bridging region, the quantum dot, 
described by discrete electronic multiplets
\bea
\hat{\mathcal{H}}=\hat{T}+\hat{V}+\hat{\mathcal{H}}_{int}\,,
\label{hint}
\eea
where $\hat{V}$ describes the tunnelling between the leads and the nanocontact and 
$\hat{\mathcal{H}}_{int}$ the local interaction in the nanocontact.

One assumes that initially the leads are not coupled through the bridging region, each lead being subject to a different 
electrochemical potential. Such a situation can be described by a density matrix
\bea
\rho_0=e^{-\beta \hat{\mathcal{H}}_0(\Phi)}/\mathrm{Tr}(e^{-\beta \hat{\mathcal{H}}_0(\Phi)})
\label{rhozeroch4}
\eea
where
\bea
\hat{\mathcal{H}}_0(\Phi) = \hat{\mathcal{H}}_0 + \Phi \hat{Y}_0\,, 
\eea
with $\mathcal{H}_0$ the non-interacting 
Hamiltonian of the independent left ($L$) and right ($R$) leads plus 
the nanocontact
\be
\hat{\mathcal{H}}_0\equiv\hat{T}+\hat{\mathcal{H}}_{int}\,,
\ee 
$\Phi$ is the applied voltage between the two leads,
and
\bea
\Phi\hat{Y}_0 = \Phi\left(\hat{N}_L-\hat{N}_R\right)/2
\eea
that describes the electrostatic energy gain due to the presence of the 
bias voltage, where $\hat{N}$ is the number operator
-- the initial 
state is stationary though out-of-equilibrium, equilibrium meant to be the two leads at the same chemical potential. 

Suddenly the coupling to the bridging region is switched on - namely 
the Hamiltonian changes from 
$\hat{T}+\hat{\mathcal{H}}_{int}$ into $\hat{\mathcal{H}}=\hat{T}+\hat{\mathcal{H}}_{int}+\hat{V}$ -
and a current starts to flow.  If 
\bea
U(t)=e^{-i\hat{\mathcal{H}}t}
\label{evol}
\eea
is the time evolution operator with the full interaction, 
the initial density matrix
$\rho_0$ evolves in time maintaining the functional form of a Boltzmann 
exponential 
\bea
\rho(t)=e^{-\beta \hat{\mathcal{H}}(t,\Phi)}/
\mathrm{Tr}(e^{-\beta \hat{\mathcal{H}}(t,\Phi)})
\label{rhotimech4}
\eea
where 
\bea
\hat{\mathcal{H}}(t,\Phi)=\hat{\mathcal{H}}(t)+\Phi \hat{Y}(t)
\eea
and 
\bea
\hat{\mathcal{H}}(t) \!\!&=&\!\! U(t) (\hat{T}+\hat{\mathcal{H}}_{int}) U(t)^\dagger,\nonumber\\
\hat{Y}(t) \!\!&=&\!\! U(t) \hat{Y}_0 U(t)^\dagger  
\label{asintUch4}
\eea

For time $t$ sufficiently large, namely after a transient time $\mathcal{T}$, 
the system reaches a steady state with constant current. 
If we are interested only in steady state properties, 
a good starting point is offered by Hershfield's results.~\cite{Her.}
He showed that the stationary state value of 
certain observables coincide with their equilibrium 
value obtained through the effective density matrix 
\bea
\rho=e^{-\beta \hat{\mathcal{H}}(\Phi)}/\mathrm{Tr}(e^{-\beta \hat{\mathcal{H}}(\Phi)}),
\label{rhointeractingch4}
\eea
with 
\bea
\hat{\mathcal{H}}(\Phi) = \hat{\mathcal{H}} + \Phi \hat{Y}, 
\label{hershhamch4}
\eea
where $\hat{Y}$ is the asymptotic time evolution of $\hat{Y}_0$
still satisfying\footnote{The physical meaning of \eqref{commutationch4} is that the steady state can be reached only when all terms 
of $\hat{Y}_0$ that do not commute with the Hamiltonian $\hat{\mathcal{H}}$ have been filtered out.}     
\begin{equation}
[\hat{\mathcal{H}},\hat{Y}]=0.\label{commutationch4}
\end{equation}
Should $\hat{Y}$ be known, steady state properties could in principle
be obtained by any equilibrium technique.

\section{The resonant-model out of equilibrium}\label{theresonantmodeloutofequilibrium} 

Let us consider the simple case of a non-interacting single-level 
quantum dot
\be
\hat{\mathcal{H}}_0=\hat{T}+\hat{V}\,,
\label{Hamfreech4}
\ee
with
\bea
\hat{T} \!\!&=&\!\! \sum_{\alpha=-1,1}\sum_{k\sigma}
\epsilon_k \,c^\dagger_{\alpha k\sigma}c^\dagga_{\alpha k\sigma}  
\,+\sum_\sigma\, \epsilon_d\,\dc_\sigma\da_\sigma,
\nonumber\\
\hat{V} \!\!&=&\!\! \sum_{\alpha=-1,1}\sum_{k\sigma}
\frac{V_k}{\sqrt{\Omega}}\, d^\dagger_{\sigma} c^\dagga_{\alpha k\sigma}+H.c.
\,,
\eea
where $c^\dagger_{\alpha k\sigma}$ creates a conduction electron on
the left ($\alpha=-1$) or right 
($\alpha=1$) lead with quantum number k and spin $\sigma$
while $d^\dagger_{\sigma}$ creates an electron in the dot with 
spin $\sigma$, and $\Omega$ is the quantization volume of the system. Notice that, quite generally only a single channel of conduction 
electrons is coupled to the impurity, so that the model can always be mapped onto two one-dimensional leads hybridized at the 
contiguous edges with an impurity. Therefore it is perfectly legitimate to regard the quantum number $k$ as one-dimensional momentum 
and $\Omega$ as the linear size of the system.   

Let us assume that our system does not have bound states. 
In this case it can be proven~\cite{Her.,Hanscatt} that 
the nonequilibrium Hamiltonian 
\be
\hat{\mathcal{H}}_0(\Phi)=\hat{\mathcal{H}}_0 + \Phi \hat{Y}_0
\ee
can be expressed as 
\bea
\hat{\mathcal{H}}_0 \!\!&=&\!\! \sum_{\alpha=-1,1}\sum_{k\sigma}
\epsilon_k\,
\psi^\dagger_{\alpha k\sigma}\,\psi^\dagga_{\alpha k\sigma},\nonumber\\
\hat{Y}_0 \!\!&=&\!\! \sum_{\alpha=-1,1}\sum_{k\sigma}
\frac{\alpha}{2} \,
\psi^\dagger_{\alpha k\sigma}\,\psi^\dagga_{\alpha k\sigma}\,;
\label{H*0ch4}
\eea
where $\psi^\dagger_{\alpha k\sigma}$
are the fermionic creation operators that generate the left ($\alpha=-1$) 
and right ($\alpha=1$) incident scattering waves
\bea
\psi^\dagger_{\alpha k\sigma}\,|0\rangle \!\!&=&\!\!
\left[1+\frac{1}{\epsilon_k-\hat{\mathcal{H}}+i\,0^+}\,\hat{V}\right]
c^\dagger_{\alpha k\sigma}\,|0\rangle
\nonumber\\ \!\!&=&\!
c^\dagger_{\alpha k\sigma}\,|0\rangle+ 
 \frac{V_k}{\sqrt{\Omega}}\, g_{d}(\epsilon_k)\, d^\dagger_{\sigma}\,|0\rangle \nonumber \\
&&\!\!\!\!\!\!\!+
\sum_{\alpha' k'\sigma'}\!\! \frac{V_kV_{k'}}{\Omega} 
\frac{g_{d}(\epsilon_k)}{\epsilon_k-\epsilon_{k'}+i\,0^+}\,
c^\dagger_{\alpha' k'\sigma'}|0\rangle ;
\label{scatt-freech4}
\eea  
being $g_{d}(\epsilon)$ the retarded Green's function of the impurity at 
equilibrium, which, in the infinite bandwidth limit, is given by
\be
g_{d}(\epsilon)=\frac{1}{\epsilon-\epsilon_d+i\Gamma}\,.
\ee  
We underline that Eq.~\eqref{scatt-freech4} is meaningful only in the 
thermodynamic limit, i.e. when $\Omega\rightarrow\infty$. 
For a finite system the time evolution of an incident
state
\be
|\psi^{in}_{\alpha k\sigma}\rangle=c^\dagger_{\alpha k\sigma}|0\rangle
\ee
oscillates, namely it doesn't converge to a well defined scattering state
\be
|\psi_{\alpha k\sigma}\rangle=\psi^\dagger_{\alpha k\sigma}|0\rangle\,.
\ee
The scattering states~\eqref{scatt-freech4} constitute, 
in the thermodynamic limit, a complete basis
\be
\sum_{\alpha k\sigma}\psi^\dagger_{\alpha k\sigma}\,\psi^\dagga_{\alpha k\sigma}
=\sum_{\alpha k\sigma}\cc_{\alpha k\sigma}\ca_{\alpha k\sigma}
+\sum_{\sigma}\dc_{\sigma}\da_{\sigma}\,,
\label{scattcompletenessch4}
\ee
provided that there exist no bound states.~\cite{Hanscatt}
Eq.~\eqref{scattcompletenessch4} allows us to formally expand 
the $c$ and $d$ operators as follows
\bea
\cc_{\bar{\alpha}\bar{k}\bar{\sigma}}\!\!&=&\!\!
\psi^\dagger_{\bar{\alpha}\bar{k}\bar{\sigma}}
+\sum_{\alpha k}\frac{V_{\bar{k}}V_k}{\Omega}\,
\frac{g^*(\epsilon_k)}{\epsilon_k-\epsilon_{\bar{k}}-i0^+}\,
\psi^\dagger_{\alpha k\bar{\sigma}}
\nonumber \\
\dc_{\bar{\sigma}} \!\!&=&\!\! 
\sum_{\alpha k}\frac{V_k}{\sqrt{\Omega}}\,g^*(\epsilon_k)\,
\psi^\dagger_{\alpha k\bar{\sigma}}\,,
\label{scatterchemiserveinversech4}
\eea  
and to calculate the average of any operator using the result 
\be
\Av{\Psi(\Phi)}{\psi^\dagger_{\alpha  k}\psi^\dagga_{\alpha' k'}}
=\delta_{\alpha\alpha'}\delta_{k k'}\,
f\left(\epsilon_k+\Phi\frac{\alpha}{2}\right)
\label{altrarobachemiserveinversech4}
\ee
- where $|\Psi(\Phi)\rangle$ is the ground state of 
$\hat{\mathcal{H}}_0(\Phi)$ and 
$f(\epsilon)$ is the Fermi function. The correct value
of the average is finally obtained taking the limit for 
$\Omega\rightarrow\infty$ of the result. It can be proven that
the obtained value is the same that one could obtain within the Keldish 
technique.

It is very important to underline that 
the scattering operators can formally be defined even in the interacting case,
although their explicit calculation is not feasible in practice.
If, for instance, we add to the resonant model Hamiltonian~\eqref{Hamfreech4}   
a Hubbard repulsion term on the impurity 
\be
\hat{U}=\frac{U}{2}(\hat{n}_d-1)^2
\ee
(Anderson impurity model), the scattering operators $\psi^\dagger_{\alpha k\sigma}$
are defined as the asymptotic time evolution of the $c^\dagger_{\alpha k\sigma}$ operators 
generated by the full Anderson Hamiltonian
\bea
\hat{\mathcal{H}}=\hat{T}+\hat{V}+\hat{U}\,,
\label{inta}
\eea  
and still satisfy
the completeness relation~\eqref{scattcompletenessch4} 
in the absence of bound states.~\cite{Hanscatt}
Moreover the interacting Hamiltonian~\eqref{inta} can still be 
expressed in terms of scattering
states~\cite{Hanscatt}
\bea
\hat{\mathcal{H}} \!\!&=&\!\! \sum_{\alpha=-1,1}\sum_{k\sigma}
\epsilon_k\,
\psi^\dagger_{\alpha k\sigma}\,\psi^\dagga_{\alpha k\sigma},\nonumber\\
\hat{Y} \!\!&=&\!\! \sum_{\alpha=-1,1}\sum_{k\sigma}
\frac{\alpha}{2} \,
\psi^\dagger_{\alpha k\sigma}\,\psi^\dagga_{\alpha k\sigma}\,.
\label{Hscch4}
\eea

We conclude this section calculating  the energy $\mathcal{E}_\Phi$
for the non-interacting model~\eqref{Hamfreech4}. More precisely, we consider
\be
\delta\mathcal{E}_\Phi=\mathcal{E}_\Phi - \mathcal{E}^0_{\Phi}\,,
\ee
where $\mathcal{E}^0_{\Phi}$ is the energy of the uncorrelated system $\hat{T}$.
Using Eq.~\eqref{scatterchemiserveinversech4} it can be proven that
\bea
\delta\mathcal{E}_\Phi \!\!&=&\!\! -T\sum_{n,\alpha}
\ln\!\left(\fract{i\epsilon_n \!+\! \alpha\frac{\Phi}{2}- \epsilon_d - \Delta\left(i\epsilon_n \!+\! \alpha\frac{\Phi}{2}\right)}{i\epsilon_n \!+\! \alpha\frac{\Phi}{2} 
- \epsilon_d}
\right)\nonumber\\
\!\!&=&\!\!- \int \frac{d\epsilon}{\pi} \,\delta(\epsilon)\left[
f\left(\epsilon+\frac{\Phi}{2}\right) + f\left(\epsilon-\frac{\Phi}{2}\right)\right]
\,;\label{coolresch4}
\eea
where $\epsilon_n$ are Matsubara frequencies,
\be
\Delta(z)=\frac{1}{\Omega}\sum_{k\alpha}\frac{V_k^2}{z-\epsilon_k}
\ee
is the hybridization function, and
\be
\delta(\epsilon) 
= -\mathrm{Im}\;\ln\left(\fract{\epsilon+i0^+ - \epsilon_d - \Delta(\epsilon+i0^+)}{\epsilon+i0^+ - \epsilon_d}
\right)\,.\label{def-phase-shift}
\ee

If we assume that the half-bandwidth $W$ is the unit of energy,
that the density of states is flat 
\bea
\Delta(z)\!\!&=&\!\! 
\int \frac{d\epsilon}{\pi}\frac{\Gamma(\epsilon)}{z-\epsilon}
\nonumber\\
\Gamma(\epsilon)\!\!&=&\!\! \Gamma\chi_{[-1,1]}(\epsilon)\nonumber\\
\chi_{[-1,1]}(\epsilon) \!\!&=&\!\!
\left\{
\begin{array}{rl}
1 & \quad\forall\epsilon\in [-1,1] \\
0 & \quad\forall\epsilon\not\in [-1,1]
\end{array}
\right.\,,
\eea
and that $\Gamma\ll W=1$, 
it can be easily verified that 
\be
\Gamma^2\,\frac{\partial}{\partial \Gamma}
\left(\frac{\delta\mathcal{E}_\Phi(\Gamma)}{\Gamma}\right)
=-\frac{2}{\pi}\,\epsilon \left.\arctan\left(\frac{\Gamma}{\epsilon}\right)
\right]^{-\frac{\Phi}{2}}_{-1}\,.
\label{diffdarisch4}
\ee
We observe that when $\Phi=0$ the right member of Eq.~\eqref{diffdarisch4}
is $2\Gamma/\pi$, so that the solution of Eq.~\eqref{diffdarisch4} is
\be
\delta\mathcal{E}_0=-\frac{2}{\pi}\,\Gamma 
\log\left(\frac{e}{\Gamma}\right)
\ee
(being $e$ the Nepero's number),
which derives from an hybridization gain
\be
\delta\mathcal{E}_{hyb}=-\frac{4}{\pi}\,\Gamma 
\log\left(\frac{1}{\Gamma}\right)
\label{hybgain}
\ee
and a bath energy cost
\be
\delta\mathcal{E}_{bath}=\frac{2}{\pi}\,\Gamma 
\log\left(\frac{1}{e\Gamma}\right)\,.
\ee

In the regime 
\be
W=1\gg\Phi\gg\Gamma\,,
\ee
the right member of Eq.~\eqref{diffdarisch4} vanishes, so that
\be
\delta\mathcal{E}_\Phi=\frac{2}{\pi}\,\Gamma 
\log\left(\frac{\Phi}{2}\right)\,.
\label{nonlocfuncspecregch4}
\ee

\section{The bias as an ``order parameter''}\label{thebiasasanorderparameter}

In this section we propose a point of view of the nonequilibrium problem in quantum dots
based on ideas and definitions very similar to those encountered in the general theoretical description
of collective phenomena in quantum mechanics.~\cite{Sewell}

In the standard quantum theory of \emph{finite systems}
there is a one-to-one correspondence between the observables and the
operators in a certain Hilbert space. This correspondence is unique (Von Neumann 1955), 
and this ensures that the choice of a specific representation of a finite system 
does not lead to loss of generality.
The situation is different for \emph{infinite systems}, because Von Neumann's theorem can
no longer be applied. The observable of an infinite systems generally admit a big variety of 
inequivalent representations, corresponding to macroscopically different classes of states. 

Let us consider, for instance, the case of the Heisenberg model
\be
\hat{\mathcal{H}}=-J\sum_n\mathbf{S}_n\mathbf{S}_{n+1}\,,\quad J>0\,.
\ee 
A ground state $|\Psi\rangle$ of the system has all the spins aligned parallel in the same
direction. 
The class of all the states obtained applying a finite number of \emph{local}
spin transformations to $|\Psi\rangle$ does not change its magnetic order parameter
\be
\mathbf{m}=\lim_{N\rightarrow\infty}
\sum_{n=-N}^N \frac{\Av{\Psi}{\mathbf{S}_n}}{2N+1}\,,
\label{magordheis}
\ee 
because the system is $\emph{infinite}$.
Furthermore the topological closure of the space spanned by these states (that is a Hilbert space) 
is the basis of an irreducible representation of the algebra of the Pauli spins $\{\mathbf{S}_n\}$. 
In this example we have a set of different \emph{phases} corresponding to different (inequivalent) 
representations of the observables for any arbitrary direction $\mathbf{m}$.
In other words, a state $|\Psi_{\mathbf{m}}\rangle$ defines a corresponding \emph{island} of states $\mathcal{I}(\Psi_{\mathbf{m}})$
that share the same order parameter and are
the basis of a specific representation of the algebra of the observables.

The general formal definition of the islands is based on the fundamental Gelfand-Naimark-Segal (GNS) 
theorem,~\cite{Sewell} that tells us that if we have an algebra 
$\mathcal{A}$ generated by the \emph{local} observables of 
an infinite system - i.e. all the operators that belong to any \emph{finite} region of space 
$\Lambda$, -
for each \emph{state} $\rho$ there is an operator-representation $\hat{\mathcal{A}_\rho}$ of $\mathcal{A}$ 
on a Hilbert space $\mathcal{H}_\rho$, which is determined (up to unitary equivalence) by the conditions that
it exists a vector $|\Psi_\rho\rangle$ in $\mathcal{H}_\rho$ such that
\be
\rho(A)=\Av{\Psi_\rho}{\hat{A}}\quad\forall A\in\mathcal{A}\,,
\ee
and that $\mathcal{H}_\rho$ is generated by applying the elements of $\hat{\mathcal{A}_\rho}$ to $|\Psi_\rho\rangle$. 

The island $\mathcal{I}(\rho)$ is, by definition, the set of all the states $\rho'$ corresponding to 
all the density matrices $\hat{\rho}'$ in $\mathcal{H}_\rho$.
Physically, the meaning of the states $\rho'\in\mathcal{I}(\rho)$ is that these are states generated 
by localized modifications of $\rho$, but are ``macroscopically'' equivalent to one other.

Let us now consider the system represented in Fig.~\ref{singledotsystem}, whose dynamics is defined by the 
interacting Anderson impurity model
\bea
\hat{\mathcal{H}} &=& \sum_{\alpha k\sigma}\epsilon_k \,c^\dagger_{\alpha k\sigma}c^\dagga_{\alpha k\sigma}  + \,\sum_{\alpha k\sigma} \frac{V_k}{\sqrt{\Omega}}\, d^\dagger_{\sigma} c^\dagga_{\alpha k\sigma}+H.c.\nonumber \\
&+& \sum_\sigma \epsilon_d d^\dagger_{\sigma} d^\dagga_{\sigma}
+ \frac{U}{2}(\hat{n}_d-1)^2\,,
\label{Hamand}
\eea
where 
\be
\hat{n}_{d}=\sum_\sigma d^\dagger_{\sigma} d^\dagga_{\sigma}
\label{CRch4tot}
\ee
is the impurity number operator.

To the initial state $\rho_0(\Phi,T)$ defined in Eq.~\eqref{rhozeroch4}
will correspond, through the GNS theorem, an \emph{island} of states 
\be
\mathcal{I}(\Phi,T)\equiv\mathcal{I}\left(\rho_0(\Phi,T)\right)\,,
\ee
and the islands obtained from initial states with different $\Phi$ will correspond to inequivalent phases,
because of the infinite volume of the two leads. In this sense we can say that  
$\Phi$ plays the same role as the magnetic order parameter in the example of the Heisenberg model
considered above.

In this work we assume that the dynamics of the system does not mix vectors belonging to different phases,
i.e. that if $\rho'$ belongs to $\mathcal{I}(\Phi,T)$ then so does $\rho'(t)$. This corresponds
to the physical idea that the system never equilibrates because the two leads are infinite and the junction between
them (the dot) is finite, so that the current through the dot
can not change the densities of the two leads defined by the value of $\Phi$.

The stability of the dynamics in $\mathcal{I}(\Phi,T)$ allows us to consider 
$\hat{\mathcal{H}}_{\Phi,T}$ the generator of the time evolution transformation in  
$\mathcal{I}(\Phi,T)$, i.e. the Hamiltonian that describes the dynamics of the island.

The basis of the ideas proposed in this paper is that an equilibrium problem corresponds to the study of the $\Phi=0$ phase,
while a nonequilibrium problem is equivalent to study a finite-$\Phi$ phase. Once the 
$\hat{\mathcal{H}}_{\Phi,T}$ operator is defined we can apply, in principle, any equilibrium technique
to study the physics of the corresponding island.

\section{Variational approach at $T=0$}\label{mainsec}

In this section we will concentrate our attention to a general GNS
representation of the system at $T=0$
\be
\mathcal{I}(\Phi)\equiv\mathcal{I}(\Phi,T=0)\,.
\ee
In this case we know that the initial state is represented by a pure vector 
\bea
|\Psi_0(\Phi)\rangle&\in&\mathcal{I}(\Phi)\nonumber\\
\rho_0(\Phi,T=0)&=&\ket{\Psi_0(\Phi)}\bra{\Psi_0(\Phi)}\,,
\eea
and that the operator that governs the dynamics of ${\mathcal{I}}(\Phi)$ 
is simply given by
\bea
\hat{\mathcal{H}}_\Phi&\equiv&\hat{\mathcal{H}}_{\Phi,T=0}\nonumber\\
\hat{\mathcal{H}}_\Phi&=&P_\Phi\hat{\mathcal{H}}P_\Phi\,,
\label{hamblockisl}
\eea
where $P_\Phi$ is the projector on $\mathcal{I}(\Phi)$. 

We define $\bar{\mathcal{I}}(\Phi)$ as the product of 
the Hilbert space of the dot and the Hilbert space 
generated by $|\Psi_0(\Phi)\rangle$ and all the states 
\be
|\Psi_{0S}\rangle=\prod_{(\alpha,k,\sigma)\in S}
c^\dagger_{\alpha k\sigma}|0\rangle
\,,
\label{scatt00}
\ee
where $S$ is any subset of
\be
E=\left\{(\alpha,k,\sigma)\,|\,\alpha=\pm 1, k\in[-\pi,\pi], \sigma=\pm 1/2\right\}
\label{Eset}
\ee
that differs from the set ${S}_\Phi$ of the occupied states of 
\be
|\Psi_0(\Phi)\rangle\equiv\prod_{(\alpha,k,\sigma)\in {S}_\Phi}
c^\dagger_{\alpha k\sigma}|0\rangle\,.
\label{defsphi}
\ee
by an arbitrary, but \emph{finite}, number of particle-hole transformations.
In other words, any state $|\Psi_{0S}\rangle$ can be written in the form
\be
|\Psi_{0S}\rangle=
\prod_{(\alpha,k,\sigma)\in\mathcal{K}_S}
\prod_{(\alpha',k'\sigma')\in\mathcal{K}'_S}
\!\!\!\!\cc_{\alpha k\sigma}\ca_{\alpha' k'\sigma'}
|\Psi_0(\Phi)\rangle\,,
\ee
where $\mathcal{K}_S$ and $\mathcal{K}'_{S}$ are \emph{finite} subsets of $E$.

We are going to prove that $\mathcal{I}(\Phi)$ coincides with 
$\bar{\mathcal{I}}(\Phi)$.
To prove that $\bar{\mathcal{I}}(\Phi)\subset{\mathcal{I}}(\Phi)$ we 
observe that whether a state $|\Psi\rangle\in\bar{\mathcal{I}}(\Phi)$ belongs 
to ${\mathcal{I}}(\Phi)$ also the state
\be
|\Psi^\Lambda_{\alpha k\sigma,\alpha' k'\sigma'}\rangle\equiv
\sum_{R R'\in\Lambda}e^{ikR}e^{-ik'R'}\cc_{\alpha R\sigma}\ca_{\alpha' R'\sigma'}|\Psi\rangle
\label{forincl1}
\ee
belongs to it for any finite region of space $\Lambda$.
But, by definition, ${\mathcal{I}}(\Phi)$ is the \emph{topological closure} of the space 
generated by the states obtained modifying $|\Psi_0(\Phi)\rangle$ locally, so that the state
\be
\cc_{\alpha k\sigma}\ca_{\alpha' k'\sigma'}|\Psi\rangle=
\lim_{\Lambda\uparrow}|\Psi^\Lambda_{\alpha k\sigma,\alpha' k'\sigma'}\rangle\,,
\label{forincl1bis}
\ee
where $\lim_{\Lambda\uparrow}$ denotes the limit for the size 
$|\Lambda|$ of $\Lambda$ going to infinity, 
belongs to ${\mathcal{I}}(\Phi)$ too. 
The inclusion $\bar{\mathcal{I}}(\Phi)\subset{\mathcal{I}}(\Phi)$
is then proven by induction.

To prove that ${\mathcal{I}}(\Phi)\subset\bar{\mathcal{I}}(\Phi)$
we consider again the set $E$ defined in Eq.~\eqref{Eset}
of all the possible values of $(\alpha,k,\sigma)$,
and we associate to each subset $e\subset E$ the operator
\be
N^\Lambda_e=\!\!\!\sum_{\left(\alpha,k=\frac{2\pi}{|\Lambda|}n,\sigma\right)\in e}
\cc_{\alpha,k,\sigma}\ca_{\alpha,k,\sigma}\quad (\,n\; \text{integer}\,)\,.
\ee

Let us consider a general state $|\Psi\rangle\in{\mathcal{I}}(\Phi)$.
If $|\Psi\rangle$ does not belong to $\bar{\mathcal{I}}(\Phi)$
then, by definition, one can define a set $e\subset E$ of 
single particle states such that either 
\bea
\lim_{\Lambda\uparrow}\frac{1}{|\Lambda|}\Av{\Psi}{N^\Lambda_e}\!\!&>&\!\!0\nonumber\\
\lim_{\Lambda\uparrow}\frac{1}{|\Lambda|}\Av{\Psi_0(\Phi)}{N^\Lambda_e}\!\!&=&\!\!0
\label{ordpar1}
\eea
or
\bea
\lim_{\Lambda\uparrow}\frac{1}{|\Lambda|}\Av{\Psi_0(\Phi)}{N^\Lambda_e}\!\!&>&\!\!0
\nonumber\\
\lim_{\Lambda\uparrow}\frac{1}{|\Lambda|}\Av{\Psi}{N^\Lambda_e}\!\!&=&\!\!0\,,
\label{ordpar2}
\eea
But, if such set exists, it is clear that $|\Psi\rangle$ can't be generated by the application 
of local observables to $|\Psi_0(\Phi)\rangle$, because the contribution
of local modifications to Eqs.~(\ref{ordpar1}-\ref{ordpar2}) vanishes in the 
limit of $|\Lambda|\rightarrow\infty$. 
The inclusion ${\mathcal{I}}(\Phi)\subset\bar{\mathcal{I}}(\Phi)$ is then proven.

We can reformulate the statement ${\mathcal{I}}(\Phi)=\bar{\mathcal{I}}(\Phi)$ 
saying that the measure $d\mu_{\Psi}(\alpha,k,\sigma)$ such that
\be
\int_e d\mu_\Psi(\alpha,k,\sigma)\equiv
\lim_{\Lambda\uparrow}\frac{1}{|\Lambda|}\Av{\Psi}{N^\Lambda_e}
\quad\forall e\subset E
\ee
is an \emph{order parameter} that identifies the \emph{phase} $\mathcal{I}(\Phi)$;
i.e. that the states $\ket{\Psi}$ that belong to ${\mathcal{I}}(\Phi)$
are characterized by the condition
\be
d\mu_{\Psi}(\alpha,k,\sigma)=d\mu_{\Psi_0(\Phi)}(\alpha,k,\sigma)\,.
\ee
We underline the strong analogy between the measure 
$d\mu_{\Psi}(\alpha,k,\sigma)$ and 
the magnetic order parameter $\mathbf{m}$, see Eq.~\eqref{magordheis}, 
for the Heisenberg model.

As we have anticipated in the previous section, we assume that
${\mathcal{I}}(\Phi)$ is stable respect to the dynamics induced by the dot. 
While such assumption is very reasonable for \emph{finite} time evolutions, 
it is less trivial that the steady state, which is reached only after an 
\emph{infinite} time, still belongs to ${\mathcal{I}}(\Phi)$ -- and 
this is what we need. 
Although we can't prove the stability of the ${\mathcal{I}}(\Phi)$ respect to 
the asymptotic dynamic induced by an interacting dot, it is encouraging 
to observe that when $U=0$
\be
d\mu_{\Psi(\Phi)}(\alpha,k,\sigma)=d\mu_{\Psi_0(\Phi)}(\alpha,k,\sigma)\,.
\ee
This can be easily verified from the following equation
\be
|\Psi(\Phi)\rangle\equiv\prod_{(\alpha,k,\sigma)\in {S}_\Phi}
\psi^\dagger_{\alpha k\sigma}|0\rangle\,,
\label{scattisl0}
\ee
where the scattering operators $\psi^\dagger_{\alpha k\sigma}$ are given by 
Eq.~\eqref{scatt-freech4} and ${S}_\Phi\subset E$ is defined by 
Eq~\eqref{defsphi}.

As a consequence of the stability of the (asymptotic) dynamics induced by the dot we can characterize ${\mathcal{I}}(\Phi)$,
in the absence of bound states 
(when Eq.~\eqref{scattcompletenessch4} 
is satisfied),~\cite{Hanscatt}
even as the space 
generated by $|\Psi(\Phi)\rangle$ and all the asymptotic time evolutions
of the eigenstates of $\hat{\mathcal{H}}_0$ defined in 
Eq.~\eqref{scatt00}, i.e. the states
\be
|\Psi_S\rangle=\prod_{(\alpha,k,\sigma)\in S}\psi^\dagger_{\alpha k\sigma}|0\rangle\,,
\label{scattisl}
\ee
where $S$ differs from the set ${S}_\Phi$ of the occupied states 
of $|\Psi(\Phi)\rangle$ by an arbitrary, but \emph{finite}, number of differences, and
$\psi^\dagger_{\alpha k\sigma}$ are the scattering operators of Eq.~\eqref{Hscch4} 
(that are interacting in general).

Starting from the above characterization of $\mathcal{I}(\Phi)$
we can understand that all the possible eigenstates $\ket{\Psi_S}$ 
of $\hat{\mathcal{H}}$ defined
in $\mathcal{I}(\Phi)$ -- i.e. the eigenstates of $\hat{\mathcal{H}}_\Phi$ 
-- correspond to the same current
\be
\label{currform}
I_S = -i\sum_{k\sigma} \frac{V_k}{\sqrt{\Omega}}
\Av{\Psi(\Phi)}{\dc_\sigma\ca_{k\sigma,-1}} \!-\! c.c.\,,
\ee
and can be considered ``equivalent'' in this sense.
Any of the eigenstates $\ket{\Psi_S}$ of $\hat{\mathcal{H}}$ 
in $\mathcal{I}(\Phi)$ defined in Eq.~\eqref{scattisl}. 
is, in fact, the asymptotic time evolution of the corresponding 
``initial'' state
\be
|\Psi_{0S}\rangle\equiv\prod_{(\alpha,k,\sigma)\in S}
c^\dagger_{\alpha k\sigma}|0\rangle\,.
\ee
The existence of a particular set  $\bar{S}$ such that 
\be
{I}_{\bar{S}}\neq I\equiv\Av{\Psi(\Phi)}{\hat{I}}
\ee
would imply that the state $|\Psi_{0\bar{S}}\rangle$ 
-- obtained, by definition,  modifying the initial state 
$\ket{\Psi_0(\Phi)}$ applying to it only a finite number of particle-hole transformation --
leads, after an infinite transient time,
to a different current respect to 
the one of the steady state $\ket{\Psi(\Phi)}$, and this is clearly unphysical
(notice that the same argument can be applied to any \emph{local} observable, and not 
only to the current operator).

It is interesting to check directly the validity of the above statement
for the simple non-interacting case $U=0$.
The scattering operators are, in this case, given by Eq.~\eqref{scatt-freech4}, and
the average of the current operator can be calculated with 
Eq.~\eqref{scatterchemiserveinversech4} and the identity
\be
\Av{\Psi_S}{\psi^\dagger_{\alpha  k\sigma}\psi^\dagga_{\alpha' k'\sigma'}}
\!=\delta_{\alpha\alpha'}\delta_{k k'}\delta_{\sigma\sigma'}\,
n_{\Psi_S}\!\left(\alpha,k,\sigma\right);
\label{altrarobagen}
\ee
where
\be
n_{\Psi_S}\!\left(\alpha,k,\sigma\right)=
\left\{
\begin{array}{rl}
1 & \text{if}\; \left(\alpha,k,\sigma\right)\in S[\Psi] \\
0 & \text{if}\; \left(\alpha,k,\sigma\right)\not\in S[\Psi]\,.
\end{array}
\right.
\ee
The result is 
\be
\label{currform0}
I =\int d\epsilon\, \Gamma(\epsilon)\rho_d(\epsilon)
\sum_{\alpha\sigma} \alpha\, n_{\Psi_S}\!\left(\alpha,\epsilon,\sigma\right)\,,
\ee
where $\rho_d(\epsilon)$ is the spectral function of the dot
\bea
\rho_d(\epsilon) =
-\frac{1}{\pi}\,\text{Im}\left(g_d(\epsilon+i0^+)\right)\,.
\eea  
If $|\Psi\rangle\in\mathcal{I}(\Phi)=\bar{\mathcal{I}}(\Phi)$ then
\be
n_{\Psi_S}\!\left(\alpha,\epsilon,\sigma\right)\simeq f\left(\epsilon+\Phi\frac{\alpha}{2}\right)\,,
\ee
i.e. the difference between the two functions does not contribute
to the integral~\eqref{currform0} that defines the value of the current $I$, because,
by definition, 
\be
\sum_{\alpha\sigma}\int d\epsilon \left|
n_{\Psi_S}\!\left(\alpha,\epsilon,\sigma\right)- f\left(\epsilon+\Phi\frac{\alpha}{2}\right)
\right|= 0\,.
\ee

In order to identify the eigenstates of $\hat{\mathcal{H}}_\Phi$ a possibility is to minimize
the variance
\bea 
\sigma^2_\Phi\left[\Omega^G_+\right]\!\!&=&\!\!
\frac{\Av{\Psi_0(\Phi)}{\Omega^{G\dagger}_+\hat{\mathcal{H}}^2_\Phi\Omega^{G}_+}}{\Av{\Psi_0(\Phi)}{\Omega^{G\dagger}_+\Omega^{G}_+}}
\nonumber\\
&&-\left(\frac{\Av{\Psi_0(\Phi)}{\Omega^{G\dagger}_+\hat{\mathcal{H}}_\Phi\Omega^{G}_+}}{\Av{\Psi_0(\Phi)}{\Omega^{G\dagger}_+\Omega^{G}_+}}\right)^2
\eea
respect to the most general operator $\Omega^{G}_+$ that is generated by the algebra of the \emph{local} observables.
Notice that what we have defined is a \emph{variational principle} for the Hershfield steady state at zero temperature!

\subsection{An energy-based approach}

One may be tempted to use the energy instead of the variance, i.e. to claim that the 
ground state can be calculated variationally even minimizing the energy
\be
\mathcal{E}_\Phi\left[\Omega^G_+\right]=
\frac{\Av{\Psi_0(\Phi)}{\Omega^{G\dagger}_+\hat{\mathcal{H}}_\Phi\Omega^{G}_+}}{\Av{\Psi_0(\Phi)}{\Omega^{G\dagger}_+\Omega^{G}_+}}
\,.
\label{enfunc}
\ee
respect to the $\Omega^{G}_+$-operators defined above.

Unfortunately such energy minimum does not exist, 
because a $\Phi>0$ phase contains states obtained from 
the Hershfield state $|\Psi(\Phi)\rangle$ moving an arbitrary (although finite) number of electrons from one of the leads to the other.
A solution to this problem would be to consider only the limited subset of $\Omega^{G}_+$ such that $\Omega^{G}_+|\Psi_0(\Phi)\rangle$ 
satisfy the equation
\be
\hat{Y}\,\Omega^{G}_+|\,\Psi_0(\Phi)\rangle=\delta N_0(\Phi)\,\Omega^{G}_+|\Psi_0(\Phi)\rangle
\label{eigencond}
\ee
where
\be
\hat{Y} = \sum_{\alpha=-1,1}\sum_{k\sigma}
\frac{\alpha}{2} \,
\psi^\dagger_{\alpha k\sigma}\,\psi^\dagga_{\alpha k\sigma}
\label{eigen}
\ee
and $\delta N_0(\Phi)$ is defined by
\be
\hat{Y}_0\,|\Psi_0(\Phi)\rangle=\delta N_0(\Phi)\,|\Psi_0(\Phi)\rangle
\label{eigen0}
\ee
where
\be
\hat{Y}_0=\sum_{\alpha=-1,1}\sum_{k\sigma}
\frac{\alpha}{2} \,c^\dagger_{\alpha k\sigma}\,c^\dagga_{\alpha k\sigma}\,.
\ee
In this subspace it is clear, if we think in terms of interacting scattering operators, that the only eigenstate
of $\hat{\mathcal{H}}_\Phi$ is the state that minimizes the energy.
Unfortunately the operator $\hat{Y}$ is not known, so that the energy-based procedure defined here can't be applied 
rigorously in practice. 

Notice that if we formally apply the exact (unknown) asymptotic time evolution operator 
$\Omega_+$ to the two members of Eq.~\eqref{eigen0} 
\be
\Omega_+\,\hat{Y}_0|\,\Psi_0(\Phi)\rangle=\delta N_0(\Phi)\,\Omega_+|\Psi_0(\Phi)\rangle
\ee
we obtain exactly the condition defined in Eq.~\eqref{eigencond}, namely that
\bea
\hat{Y}\Omega_+\,|\,\Psi_0(\Phi)\rangle \!\!&=&\!\! \left(\Omega_+\,\hat{Y}_0\,\Omega_+^\dagger\right)\Omega_+|\Psi_0(\Phi)\rangle
\nonumber\\
\!\!&=&\!\! \delta N_0(\Phi)\Omega_+\,|\Psi_0(\Phi)\rangle\,,
\eea
where we have used that
\be
\hat{Y}=\Omega_+\,\hat{Y}_0\,\Omega^{\dagger}_+\,.
\ee
If we apply the approximated trial $\Omega^G_+$ to Eq.~\eqref{eigen0} we obtain, instead, that
\be
\Omega^G_+\,\hat{Y}_0\,|\Psi_0(\Phi)\rangle=\delta N_0(\Phi)\,\Omega^G_+\,|\Psi_0(\Phi)\rangle\,,
\ee
which is equivalent to
\be
\hat{Y}_G\,\Omega^G_+\,|\Psi_0(\Phi)\rangle=\delta N_0(\Phi)\,\Omega^G_+\,|\Psi_0(\Phi)\rangle\,,
\label{mfcond}
\ee
where
\be
\hat{Y}_G=\Omega^G_+\,\hat{Y}_0\,\Omega^{G\,-1}_+\,.
\ee
For this reason, whether we believe, for some physical reason, that the proposed $\Omega^G_+$ is sufficiently good
to guarantee (approximately)  the equivalence of Eq.~\eqref{mfcond} and Eq.~\eqref{eigencond}
the energy minimization procedure is still meaningful, although not purely variational.

In the following sections we will propose an example of such energy-based approach starting from a 
particular variational space, that is expected to describe sufficiently well the qualitative behaviour, 
in the Fermi-Liquid regime, of the interacting single-orbital Anderson impurity model.

We conclude this section observing that, at least formally, the GNS theorem mentioned above allows us to 
define a variational principle for the Hershfield steady state even at finite temperature.
In fact, to a thermal state $\rho_0(\Phi,T)$ corresponds, in its GNS representation, a vector $|\Psi_0(\Phi,T)\rangle$, 
and the dynamics of the corresponding island $\mathcal{I}(\Phi,T)$ is (presumably) still generated by some 
operator $\hat{\mathcal{H}}_{\Phi,T}$, although it is not simply a ``block'' of 
$\hat{\mathcal{H}}$ (Eq.~\eqref{hamblockisl}).  
An eigenstate of $\hat{\mathcal{H}}_{\Phi,T}$ is a stationary state, i.e. the representation of the Hershfield 
state in $\mathcal{I}(\Phi,T)$, and can be formally identified by the minimum-variance condition as before.

\section {The concept of quasi-particles out of equilibrium}\label{theconceptofquasiparticlesoutofequilibrium}

Let us consider again the general interacting system described by the Hamiltonian
\bea
\hat{\mathcal{H}}=\hat{\mathcal{H}}_0+\hat{V}+\hat{\mathcal{H}}_{int} ,
\eea
We know that if we prepare the two leads at a different chemical potentials
and let it evolve within the interacting Hamiltonian
\be
U(t)=e^{-i\hat{\mathcal{H}}t},
\ee
for times $t$ longer than some transient time $\mathcal{T}$ the final nonequilibrium
state is described by the Hershfield Hamiltonian
\bea
\hat{\mathcal{H}}(\Phi) = \hat{\mathcal{H}} + \Phi \hat{Y}, 
\eea
formally defined in Eq.~\eqref{asintUch4}

In general $\hat{Y}$ is a complicated many body operator that must satisfy Eq.~\eqref{commutationch4} and in addition 
share the same symmetry properties as $\hat{Y}_0$, i.e. a spin-singlet operator odd under interchanging the two leads. 
Therefore, generally the steady-state Hamiltonian $\hat{\mathcal{H}}(\Phi)$ is an interacting one, the interaction 
\be
\delta\mathcal{H}(\Phi)\equiv\hat{\mathcal{H}}_{int}+\Phi\left(\hat{Y}-\hat{Y}_0\right)
\ee
presumably 
remaining ``local'' (in the sense defined in section~\ref{mainsec}) as it was originally. 
Furthermore, since the 
nanocontact can not change the bulk properties of the leads, e.g. inducing a spontaneous symmetry breaking, 
$\hat{\mathcal{H}}(\Phi)$ should still describe a metal. It is therefore tempting to assume that, if in the absence of external bias 
the system, leads plus 
nanocontact, is described by a local Fermi liquid 
theory 
in the Nozi\`eres sense,~\cite{Nozieres-Journal-of-Low-Temperature-Physics}
which is generally the case, the same should hold even in the steady state after the bias is applied.~\cite{0953-8984-16-16-R01}
It then follows that it should be 
possible to represent the low energy/temperature/bias properties in terms of weakly interacting {\sl quasi-particles} 
which, by continuity 
with the non-interacting case, should be better regarded as {\sl renormalized} scattering states with an Hamiltonian of the same form 
as~\eqref{H*0ch4} with renormalized (bias dependent) energies plus additional 
weak local-interaction terms.~\cite{Nozieres-Journal-of-Low-Temperature-Physics} 
This local Fermi-liquid assumption seems to us quite plausible. However, since the bias is coupled to a non-conserved quantity, the 
charge difference between the leads, the effective bias felt by the quasi-particles will generally differ from the applied one and the 
quasi-particle current does not correspond to the real one. This implies that the current can not be expressed simply in terms of 
Landau parameters and an explicit calculation is required.

\section{The Gutzwiller approximation at equilibrium}\label{geqsec}

Let us consider the $\Phi=0$ phase, namely the equilibrium problem. Although the method we shall present 
is quite general, for sake of simplicity we shall show how it works in the simple case of a bridging region described by 
a single-orbital Anderson impurity model \emph{at half-filling}
\bea
\hat{\mathcal{H}} &=& \sum_{\alpha k\sigma}\epsilon_k \,c^\dagger_{\alpha k\sigma}c^\dagga_{\alpha k\sigma}  + \,\sum_{\alpha k\sigma} \frac{V_k}{\sqrt{\Omega}}\left( d^\dagger_{\sigma} c^\dagga_{\alpha k\sigma}+H.c.\right)\nonumber \\
&+&  \frac{U}{2}(\hat{n}_d-1)^2
\,\equiv\, \left(\hat{T}+\hat{V}\right)+\hat{U} \,\equiv\, \hat{\mathcal{H}}_0+\hat{U}
\label{Hamch4}
\eea
\begin{figure}[t!]
  \begin{center}
    \includegraphics[width=9cm]{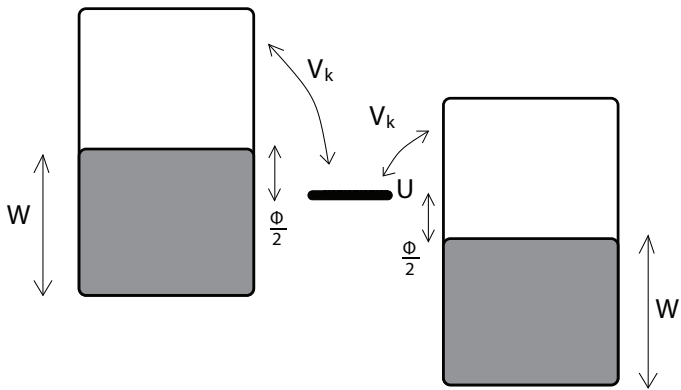}
    \caption{\label{singledotsystem}(Color online) 
      The single dot system.
    }
  \end{center}
\end{figure}

The physical properties of the above 
Anderson impurity model are very well known.~\cite{Hewson} For large $U$ the model effectively maps into a Kondo model, 
the impurity electron behaving as a local moment Kondo screened by the conduction electrons. A simple way to describe qualitatively 
and to some extent also quantitatively the Kondo screening is by a Gutzwiller-type of variational 
wavefunction~\cite{Fazekas,Fazekas1}
\be
|\Psi\rangle = \mathcal{P}_d\,|\Psi_0\rangle
\label{GWFch4}
\ee
where $\mathcal{P}_d$ is an operator that modifies the relative 
weights of the \emph{impurity} electronic configurations with respect to the 
uncorrelated wavefunction $|\Psi_0\rangle$
\bea
\mathcal{P}_d&=&\lambda_{0,2}\left(|0\rangle\langle 0|
+|\uparrow\downarrow\rangle\langle\uparrow\downarrow|\right)
\nonumber\\
&+&\lambda_{1}\left(|\uparrow\rangle\langle\uparrow|
+|\downarrow\rangle\langle\downarrow|\right)\,,
\label{explproj}
\eea
and $|\Psi_0\rangle$ is the ground state of a non-interacting variational 
resonant level Hamiltonian.

\subsection{Energy optimization}

The variational procedure amounts to optimize both the local projector $\mathcal{P}_d$ 
as well as the non-interacting wavefunction  
$\ket{\Psi_0}$ by minimizing the expectation value 
of the Hamiltonian \eqref{Hamch4}. 

We assume that $\mathcal{P}_d$ is subject to the following two 
conditions
\bea
\Av{\Psi_0}{\mathcal{P}^\dagger_d\,\mathcal{P}^\dagga_d} &=& 1,\label{unoch4}\\
\Av{\Psi_0}{\mathcal{P}^\dagger_d\,\mathcal{P}^\dagga_d\,
\hat{n}_{d\sigma}} &=& 
\Av{\Psi_0}{\hat{n}_{d\sigma}}\,,\label{duech4}
\eea 
where 
\be
\hat{n}_{d\sigma}=d^\dagger_{\sigma} d^\dagga_{\sigma}\,.
\label{CRch4}
\ee
Condition~\eqref{unoch4} is the normalization requirement
of the variational wavefunction, that corresponds, in terms of $\lambda$-parameters, to the condition
\be
\lambda_{0,2}^2+\lambda_{1}^2=2\,.
\label{normlam}
\ee
Condition~\eqref{duech4} - that ensures that all the 
Wick contractions between the conduction electron operators and the impurity 
operators are zero - allows to evaluate expectation values straightforwardly. 

In particular, the expectation value of the Hamiltonian \eqref{Hamch4}, 
that has to be minimized, is 
\begin{widetext}
\bea
E[\Psi] &=& \fract{\Av{\Psi}{\hat{\mathcal{H}}}}{\langle \Psi|\Psi\rangle} 
\nonumber\\&=& 
\langle \Psi_0 |\,\Bigg[ \sum_{\alpha k\sigma}\epsilon_k \,c^\dagger_{\alpha k\sigma}c^\dagga_{\alpha k\sigma} 
+ \,\sum_{\alpha k\sigma} \frac{R\,V_k}{\sqrt{\Omega}}\left( d^\dagger_{\sigma} c^\dagga_{\alpha k\sigma}+H.c.\right)\nonumber \Bigg]\,|\Psi_0\rangle   
+ \frac{U}{2}\,\langle \Psi_0 |\mathcal{P}^\dagger_d(\hat{n}_d-1)^2\mathcal{P}^\dagga_d|\Psi_0\rangle\nonumber \\\nonumber \\
&\equiv&
\langle \Psi_0 |
\hat{\mathcal{H}}^0_R
|\Psi_0\rangle  
+\frac{U}{2}\,\langle \Psi_0 |\mathcal{P}^\dagger_d(\hat{n}_d-1)^2\mathcal{P}^\dagga_d|\Psi_0\rangle
\label{E-varch4}
\eea  
\end{widetext}
where the hopping renormalization coefficient $R$ is obtained through the 
following equation:
\bea
&&\langle \Psi_0 |\,\mathcal{P}^\dagger_d\, d^\dagger_{\sigma}\,\mathcal{P}^\dagga_d\,
d^\dagga_{\sigma}\,|\Psi_0\rangle = 
R\,\langle \Psi_0 | d^\dagger_{\sigma}d^\dagga_{\sigma}|\Psi_0\rangle\,,
\label{Z-cfch4} 
\eea
whose solution is 
\be
R=\lambda_{0,2}\lambda_{1}\,.
\label{relRhrin}
\ee
The calculation of the first term in Eq.~\eqref{E-varch4}
reduces, provided eqs.~(\ref{unoch4}) and (\ref{duech4}) are satisfied, to 
calculate the energy gain of $\hat{\mathcal{H}}^0_R$ due to the renormalized 
tunnelling term
\bea
\hat{V}_R &=&
\sum_{\alpha k\sigma} \frac{R\,V_k}{\sqrt{\Omega}}\, d^\dagger_{\sigma} c^\dagga_{\alpha k\sigma}+H.c.\,.
\label{zero-bias-hopp-tunnch4}
\eea

The variational Hamiltonian whose ground state is the uncorrelated wavefunction $|\Psi_0\rangle$ has rigorously no physical 
meaning but for the ground state properties. However, it is common ~\cite{BunemannJorgGebhardFlorianandThul} 
to interpret it as the Hamiltonian of the quasi-particles and
\bea
R^2=z
\label{gutzqprelch4}
\eea  
as the quasi-particle weight of a single-particle excitation. 
Within such an assumption, the Gutzwiller approximation technique can be 
regarded as a tool to extract quasi-particle properties.

From now on the unit of energy is given by the conduction electron 
half-bandwidth $W$. The explicit value of the variational energy (referred to the ground state 
energy of the unperturbed system), is then given by
\be
\delta E[\Psi]=\frac{2}{\pi}z\Gamma \log\left(\frac{z\Gamma}{e}\right) + 
 \frac{U}{4}\left(1-\sqrt{1-z}\right)\,.
\label{vareneqforreferee}
\ee

\begin{figure}[t!]
  \begin{center}
    \includegraphics[width=8.0cm]{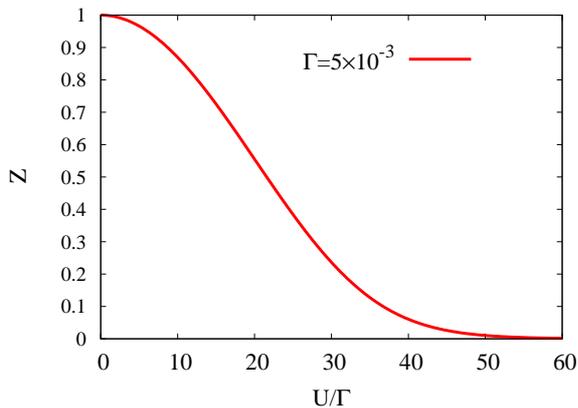}
    \caption{\label{ZUplot} (Color online) 
      $z$, as a function of $U/\Gamma$. 
    } 
  \end{center}
\end{figure}

In Fig.~\ref{ZUplot} we show the value of the optimal $z$, as a function of $U$.
At $U=0$ we find that $z=1$ (as expected) and has 
a finite curvature.
When $U\rightarrow\infty$ we find that 
\be
z(U)\sim\frac{1}{\Gamma}
\exp\left(-\frac{\pi}{16}\frac{U}{\Gamma}\right)\,.
\label{forrefereecomm1}
\ee

Notice that at large $U$ the value
of $z$ vanishes exponentially but remains finite because 
\be
\langle \Psi_0 |
\hat{\mathcal{H}}^0_R
|\Psi_0\rangle
=-\frac{2}{\pi}\,z\Gamma 
\log\left(\frac{e}{z\Gamma}\right)
\,,
\label{forrefereecomm2}
\ee
which vanishes at $z=0$ with an infinite derivative due to the
presence of $z$ in the logarithm.

We conclude this section by underlining a limit of  
the Gutzwiller method from a quantitative point of view.
Using Eq.~\eqref{forrefereecomm1} we can define the 
``Gutzwiller approximation'' for the Kondo temperature as
\be
T^G_K\,\sim z(U)\stackrel{U\gg\Gamma}{\sim}\frac{1}{\Gamma}
\exp\left(-\frac{\pi}{16}\frac{U}{\Gamma}\right)\,.
\label{tkG}
\ee
Notice that $T^G_K$ differs with respect to the correct value of 
the Kondo temperature
\be
T_K\sim\exp\left(-\frac{\pi}{8}\frac{U}{\Gamma}\right)
\ee
because:
\begin{itemize}
\item the universal prefactor in the exponent should be $\pi/8$ and not 
$\pi/16$,
\item the factor $W/\Gamma$ (which is equal to $1/\Gamma$ in our units)
in Eq.~\eqref{tkG} diverges in the infinite bandwidth limit.
\end{itemize}
The divergence of the right member of Eq.~\eqref{tkG}
for $W/\Gamma\rightarrow\infty$
reflects the unreliability of the method in this limit.
In fact, when $W/\Gamma\rightarrow\infty$,   
the Gutzwiller approximation predicts that $z\rightarrow 1$
even if $U>\Gamma$, as can be verified
directly from Eq.~\eqref{vareneqforreferee}.

\subsection{Variance optimization}\label{varoptsubsec}

For the equilibrium ground state of a system it is possible to define 
an infinite number of functionals with the same minimum.
For example, the functionals
\be
F_\mu[\Psi]=E[\Psi]+\mu\,\sigma_{\mathcal{H}}[\Psi]\,,
\label{eqfunc}
\ee
being $\sigma_{\mathcal{H}}$ the variance 
\be
\sigma_{\mathcal{H}}[\psi]=\left[
\Av{\psi}{\mathcal{H}^2}-\Av{\psi}{\mathcal{H}}^2
\right]^{\frac{1}{2}} \,,
\ee
are equivalent $\forall\mu>0$. 
Nevertheless the result obtained 
when the functional $F_\mu$ is minimized on a particular variational space
\emph{depends} on the factor $\mu$, and the choice has to be motivated on 
the basis of the specific problem considered.

In particular, if one think that his variational function is a good 
approximation of the ground state (and not of the excited states) 
of the system, instead of minimizing the energy $E$, it is sometime 
convenient to minimize the variance $\sigma_{\mathcal{H}}$. 
From the equilibrium variational 
principle, the smaller the energy is the better
the variational state will be, but, without an exact solution, it is 
hard to judge how accurate the
variational approximation is. On the contrary, the variance is very useful, 
because the smallest
possible variance, equal to zero, is known a priori, and in this case the 
variational state represent an exact eigenstate of the Hamiltonian.

We observe that in the Gutzwiller energy-minimization procedure discussed in the previous section
the optimization scheme leads to a correspondence between 
the $\lambda$-parameters (defining $\mathcal{P}_d$ through Eq.~\eqref{explproj}) and the uncorrelated wavefunction $\ket{\Psi_0}$:
once the $\lambda$-parameters were defined the corresponding $\ket{\Psi_0}$ was the ground state of a renormalized 
``variational Hamiltonian'' $\mathcal{H}^0_R $,
with $R$ given by Eq.~\eqref{relRhrin}. 
In the appendix we show the calculation of the variance assuming such correspondence.

If, for simplicity, we assume that
\bea
&&V_k=V_{k'}\quad\forall k,k'\nonumber\\
&&\Gamma\ll W=1\,,
\eea
the variance is given by.
\bea
\sigma^2_{\mathcal{H}}\left[\Psi\right]
&=&\left(1-z\right)
\left(\frac{\Gamma}{2}+\frac{12}{\pi^2}\,\Gamma^2 z \log^2\left(z^2\Gamma^2\right)\right)\nonumber\\
&+&\frac{U}{\pi}\, z \sqrt{1-z}\,
\Gamma \log\left(z^2\Gamma^2\right)+\frac{z U^2}{16}
\,.
\label{equationvariance2}
\eea

Unfortunately, the minimization of the variance functional~\eqref{equationvariance2} respect to 
the allowed values of $z$ does not lead to a physically reasonable result (not shown). 
Our conclusion is that the minimization of the variance requires, to be effective, a more 
realistic trial state respect to the simple Gutzwiller-type wavefunction~\eqref{GWFch4}
considered here. 

Following the derivation of Eq.~\eqref{equationvariance2} in the appendix
it is clear that only for Gutzwiller's wavefunction such that the 
correspondence~\eqref{relRhrin} between $|\Psi_{0}\rangle$ and $\mathcal{P}_d$  is verified
the variance is finite. 
We underline that \emph{at half-filling} such correspondence eliminates the diverging terms 
(for our system) even out of equilibrium, i.e. when $|\Psi_0\rangle$ is the ground state of 
\bea
\hat{\mathcal{H}}^{0}_{R}(\Phi) &=& 
\sum_{\alpha k\sigma}\epsilon_k \,\psi^\dagger_{\alpha k\sigma}(R)\,\psi^\dagga_{\alpha k\sigma}(R)+
\nonumber \\
&+&\, \Phi\,
\sum_{\alpha k\sigma}\frac{\alpha}{2} \,\psi^\dagger_{\alpha k\sigma}(R)\,\psi^\dagga_{\alpha k\sigma}(R)
\label{H-rin-scatt-bias-diffch4}
\eea 
instead of the ground state of
\bea
\hat{\mathcal{H}}^0_{R} \!\!&=&\!\! \sum_{\alpha k\sigma}\epsilon_k \,c^\dagger_{\alpha k\sigma}c^\dagga_{\alpha k\sigma} 
+ \sum_{\alpha k\sigma} \frac{R\,V_k}{\sqrt{\Omega}}\, d^\dagger_{\sigma} c^\dagga_{\alpha k\sigma}+H.c.
\nonumber\\ \!\!&\equiv&\!\!
\sum_{\alpha k\sigma}\epsilon_k \,\psi^\dagger_{\alpha k\sigma}(R)\,\psi^\dagga_{\alpha k\sigma}(R)
\,;
\eea  
being
\bea
\psi^\dagger_{\alpha k\sigma}(R) \!\!&=&\!\!
c^\dagger_{\alpha k\sigma}+ 
 \frac{R V_k}{\sqrt{\Omega}}\, g^R_{d}(\epsilon_k)\, d^\dagger_{\sigma} \nonumber \\
&&\!\!\!\!\!\!\!+
\sum_{\alpha' k'\sigma'}\!\! \frac{R^2\,V_kV_{k'}}{\Omega} 
\frac{g^R_{d}(\epsilon_k)}{\epsilon_k-\epsilon_{k'}+i\,0^+}\,
c^\dagger_{\alpha' k'\sigma'}
\label{scatt-freech4-rin}
\eea  
the appropriate renormalized scattering operators identified by $R$.

This observation suggests that at half-filling
the simple form of the trial function~\eqref{GWFch4}
is a reasonable variational representation not only for the ground 
state of our system, 
but also for its nonequilibrium Hershfield steady states. 
This observation relates with the Fermi-liquid assumption formulated
in Sec.\ref{theconceptofquasiparticlesoutofequilibrium}.

\section{The Gutzwiller approximation out of equilibrium}\label{thegutzwillerapproximationoutofequilibrium}

We study now the half-filled Anderson model 
\bea
\hat{\mathcal{H}} &=& \sum_{\alpha k\sigma}\epsilon_k \,c^\dagger_{\alpha k\sigma}c^\dagga_{\alpha k\sigma}  
+ \,\sum_{\alpha k\sigma} \frac{V_k}{\sqrt{\Omega}}\, d^\dagger_{\sigma} c^\dagga_{\alpha k\sigma}+H.c.\nonumber \\
&+& \frac{U}{2}(\hat{n}_d-1)^2
\eea
when it is driven out of equilibrium preparing the leads 
at two different chemical potentials (Fig.~\ref{singledotsystem}). 
Turning on the tunnelling interaction we know that a current starts to
flow and the system, after a transient time, reaches the steady state formally 
defined by Eq.~\eqref{rhointeractingch4}. At zero temperature
the steady state is therefore the ground state of 
\bea
\hat{\mathcal{H}}(\Phi) = \hat{\mathcal{H}} + \Phi \hat{Y}\,.
\eea

We want to approximate the Hershfield steady state with the usual 
equilibrium Gutzwiller variational wavefunction
\be
|\Psi\rangle = \mathcal{P}_d\,|\Psi_0\rangle
\label{varne}
\ee
which satisfies conditions~\eqref{unoch4} and~\eqref{duech4}.

The average on $|\Psi\rangle$ of the non-interacting part of $\hat{\mathcal{H}}$ 
\be
\hat{\mathcal{H}}_0 = \sum_{\alpha k\sigma}\epsilon_k \,c^\dagger_{\alpha k\sigma}c^\dagga_{\alpha k\sigma}  + \,\sum_{\alpha k\sigma} \frac{V_k}{\sqrt{\Omega}}\, d^\dagger_{\sigma} c^\dagga_{\alpha k\sigma}+H.c.
\ee
is equal to the average on $|\Psi_0\rangle$ of the renormalized non-interacting Hamiltonian 
\bea
\hat{\mathcal{H}}^0_{R} = \sum_{\alpha k\sigma}\epsilon_k \,c^\dagger_{\alpha k\sigma}c^\dagga_{\alpha k\sigma} 
+ \sum_{\alpha k\sigma} \frac{R\,V_k}{\sqrt{\Omega}}\, d^\dagger_{\sigma} c^\dagga_{\alpha k\sigma}+H.c.\,.
\label{H-rinch4}
\eea  
The difference between the equilibrium and the nonequilibrium case is that in the presence of a bias 
we can consider only Slater determinants belonging to $\mathcal{I}(\Phi)$.
In order to guarantee the (approximated) equivalence of Eq.~\eqref{mfcond} and Eq.~\eqref{eigencond}
the only reasonable state is the Hershfield steady state of 
the renormalized uncorrelated system~\eqref{H-rinch4}, namely the ground state of the Hamiltonian
\be
\hat{\mathcal{H}}^{0}_{R}(\Phi) = \hat{\mathcal{H}}^0_{R}+\Phi \hat{Y}^0_{R}
\label{H-rin-scatt-bias-diffch4-2}
\ee 
defined in Eq.~\eqref{H-rin-scatt-bias-diffch4}.
In fact, just like at equilibrium, we expect that the Hamiltonian~\eqref{H-rin-scatt-bias-diffch4} 
should describe weakly interacting quasiparticles in the presence of a bias. Our particular choice
for the variational Slater determinant $|\Psi_0\rangle$ in Eq.~\eqref{varne} simply means that the number of 
left and right quasiparticle is equal to the number of left and right particles in the unperturbed system.

Summarizing, our variational choice corresponds to approximate the asymptotic time evolution operator 
$\Omega_+$ with
\be
\Omega_+^G=\mathcal{P}_d\,\Omega_{+R}^G\,,\label{varchoicenoneq}
\ee
being $\Omega_{+R}^G$ the unitary operator (to be determined variationally) such that
\be
\Omega_{+R}^G\,c^\dagger_{\alpha k\sigma}\,\Omega_{+R}^{G\dagger}=\psi^\dagger_{\alpha k\sigma}(R)\quad\forall \alpha, k,\sigma\,.
\ee
The corresponding procedure amounts to minimize the following energy functional
\bea
&&E_\Phi[\Psi] = 
\langle \Psi_0(\Phi) 
|\,\hat{\mathcal{H}}^0_{R}\,|
\Psi_0(\Phi) \,\rangle \nonumber \\
&&+\, 
\frac{U}{2}\,\langle\Psi_0(\Phi) 
|\,\mathcal{P}^\dagger_d(\hat{n}_d-1)^2\mathcal{P}^\dagga_d\,|
\Psi_0(\Phi)\rangle \,,
\label{E-var-noneqch4}
\eea  
where $|\Psi_0(\Phi)\rangle$ is the ground state of $\hat{\mathcal{H}}^{0}_{R}(\Phi)$
(that satisfy the conditions~(\ref{unoch4}-\ref{duech4})). 
In other words, the only difference respect to the equilibrium calculation
is that we substitute the equilibrium energy gain due to the tunnelling term 
\eqref{zero-bias-hopp-tunnch4}
for the energy gain due to the tunnelling term in the non equilibrium
quasi-particle Hamiltonian \eqref{H-rin-scatt-bias-diffch4}. 
It can be easily proven that the value of the variational energy (referred to the nonequilibrium 
energy of the unperturbed system), is given by
\be
\delta E_\Phi[\Psi] = \delta \mathcal{E}_\Phi[\Psi]
+ \frac{U}{4}\left(1-\sqrt{1-z}\right)\,,
\ee
where
\bea
\delta \mathcal{E}_\Phi[\Psi]&=&-\frac{2}{\pi}\,\arctan(z\Gamma)
+\frac{\Phi}{\pi}\,\arctan\left(\frac{z\Gamma}{\Phi/2}\right)\nonumber\\
&&+\,\frac{z\Gamma}{\pi}\log\left(\frac{z^2\Gamma^2+(\Phi/2)^2}{1+z^2\Gamma^2}\right)
\eea

\begin{figure}[t!]
  \begin{center}
    \includegraphics[width=8.cm]{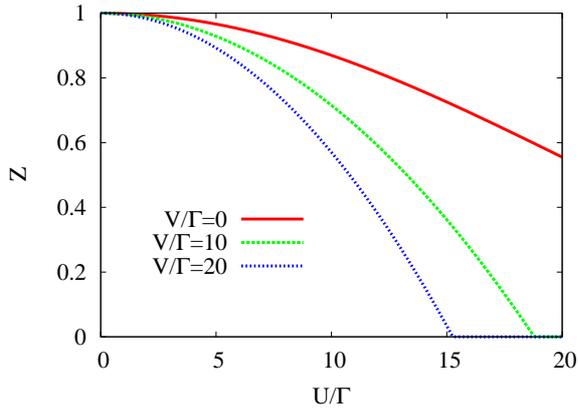}
    \caption{\label{phidepofz}(Color online) 
      $z$ as a function of $U/\Gamma$ for 
      $\Gamma=5\!\times\! 10^{-3}$ and three 
      different values of the bias $V/\Gamma$.
    }
  \end{center}
\end{figure}

We stress that our functional, and then the value of $R$ after the 
optimization, depends on the bias $\Phi$. This is crucial
in order to properly take into account the strong correlation effects 
induced by the Hubbard repulsion, i.e. to obtain the expected destruction of the 
Kondo resonance at finite bias (see Fig.~\ref{phidepofz}).

The expression for the average of the current after the optimization is
\bea
\label{currformulach4}
I &=& -i\sum_{k\sigma} \frac{V_k}{\sqrt{\Omega}}\left(
\Av{\Psi_0(\Phi)}{\dc_\sigma\ca_{k\sigma,-1}} - c.c. \right)
\nonumber \\
&=&  \int^{\frac{\Phi}{2}}_{-\frac{\Phi}{2}} d\epsilon\, \Gamma_R(\epsilon)\,\rho_d^{\Gamma_R}(\epsilon)
\eea  
where $\rho_d^{\Gamma_R}(\epsilon)$ is the spectral function of the dot, that is
\bea
\rho_d^{\Gamma_R}(\epsilon) =
\frac{1}{\pi}\frac{\Gamma_R}
{\epsilon^2+\Gamma_R^{2}}\,\chi_{[-1,1]}(\epsilon)
\eea  
with
\bea
\Gamma_R(\epsilon) &=& R^2\, \Gamma(\epsilon)
\eea  
having assumed that the
density of states is flat and that $\Gamma\ll W=1$
\bea
\Delta(z)\!\!&=&\!\! 
\int \frac{d\epsilon}{\pi}\frac{\Gamma(\epsilon)}{z-\epsilon}
\nonumber\\
\Gamma(\epsilon)\!\!&=&\!\! \Gamma\chi_{[-1,1]}(\epsilon)
\eea

We notice that Eq.~\eqref{currformulach4} fails to describe the system accurately when $\Phi\sim U$, because it doesn't 
take into account the spectral contribution of the Hubbard bands. 
However, for the simple 
single-band Anderson
model we can reproduce artificially the correct qualitative behaviour of 
the current in this regime by substituting $R^2\rho_d^{\Gamma_R}(\epsilon)$ with
\bea
\rho^U_d(\epsilon) =R^2\rho_d^{\Gamma_R}(\epsilon)+
\frac{1}{2}(1-R^2)\!\!\!\!\sum_{\alpha=-1,1}\!\! \rho_d^{\Gamma}\!
\left(\epsilon-U\frac{\alpha}{2}\right)
\eea  
in Eq.~\eqref{currformulach4}.

\begin{figure}[t!]
  \begin{center}
    \includegraphics[width=8.0cm]{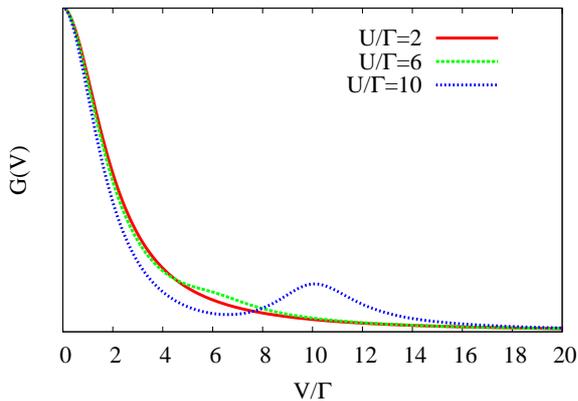}
    \caption{\label{conductance}(Color online) 
      Conductance as a function of the bias $V/\Gamma$ for 
      $\Gamma=5\!\times\! 10^{-3}$ 
      and three different values of $U/\Gamma$.
    }
  \end{center}
\end{figure}

In Fig.~\ref{conductance} we show the results for the conductance $G$ 
of the Anderson model.  
The obtained value of the conductance at zero bias is universal 
as expected, and the curvature is given by
\be
\left.\frac{d^2G}{d\Phi^2}\right|_{\Phi=0}=
-\frac{1}{2\pi(R^2\Gamma)^{2}}\sim -\frac{1}{(T^{G}_K)^2}
\ee
- $T^G_K$ being the Kondo temperature with the incorrect prefactor
predicted by the Gutzwiller method
\be
T^G_K\,\sim\,e^{-\frac{\pi}{16}\frac{U}{\Gamma}}\,.
\ee
Nevertheless for large enough value of $U$ we found 
(not shown) that the conductance may become negative, which is 
unrealistic.
In order to establish the regime of validity of our method, 
we note that the Fermi-liquid description that we assume
is applicable only for values 
of the bias much lower then the Kondo temperature $T_K$.
For the single-orbital Anderson impurity model we can calculate 
analytically the minimum value of the energy 
functional~\eqref{E-var-noneqch4} when 
\be
W\gg\Phi\gg\Gamma\,,
\ee
namely when Eq.~\eqref{nonlocfuncspecregch4} can be applied, so that 
\be
\langle \Psi_0(\Phi) 
|\,\hat{\mathcal{H}}^0_{R}+\Phi Y^0_{R}\,|
\Psi_0(\Phi) \,\rangle
=\frac{2}{\pi}\,R^2\,\Gamma 
\log\left(\frac{\Phi}{2}\right)\,.
\ee
In particular, it can be easily proven that the value of $z$ 
vanishes at
\be
\frac{\Phi^*}{2}=e^{-\frac{\pi}{16}\frac{U}{\Gamma}}\,\sim T^G_K\,,
\ee
that is out of the expected regime of validity of the calculation.

\subsection{Why is half-filling special?}

Let us consider the Anderson model away from particle-hole symmetry
\bea
\hat{\mathcal{H}} &=& \sum_{\alpha k\sigma}\epsilon_k \,c^\dagger_{\alpha k\sigma}c^\dagga_{\alpha k\sigma}  + \,\sum_{\alpha k\sigma} \frac{V_k}{\sqrt{\Omega}}\, d^\dagger_{\sigma} c^\dagga_{\alpha k\sigma}+H.c.\nonumber \\
&+& \epsilon_d\sum_\sigma\dc_\sigma\da_\sigma + \frac{U}{2}(n_d-1)^2
\eea
by requiring $\epsilon_d\neq 0$. 
The state $|\psi_0\rangle$ which minimize the
energy $\hat{\mathcal{H}^0_R}$ and satisfies Eq.~(\ref{duech4})
\be
\Av{\Psi_0}{\dc_\sigma\da_\sigma}=n
\label{condcondconddddd}
\ee
can be calculated within the Lagrange multipliers method, namely
$|\psi_0\rangle$ is the ground state of the Hamiltonian
\be
\hat{\mathcal{H}}^{\mu}_R=\hat{\mathcal{H}}^0_R+\mu\sum_\sigma(\dc_\sigma\da_\sigma-n)
\label{parappappa}
\ee
with a proper chemical potential $\mu$. 

In particular, when $\epsilon_d=0$ the ground state of 
$\hat{\mathcal{H}}^0_R$ satisfies the constraint~(\ref{duech4}) automatically,
namely $\hat{\mathcal{H}}^{\mu}_R=\hat{\mathcal{H}}^0_R$, and the corresponding
non-equilibrium Hamiltonian $\hat{\mathcal{H}}^0_R(\Phi)$
automatically satisfies the constraint~(\ref{duech4}) too,
\be
\Av{\Psi_0}{\dc_\sigma\da_\sigma}=
\Av{\Psi_0(\Phi)}{\dc_\sigma\da_\sigma}=\,\frac{1}{2}
\ee

Let us now consider the general case $\epsilon_d\neq 0$. 
In this case  
\be
\hat{\mathcal{H}}^{\mu}_R=
\sum_{\alpha k\sigma}\epsilon_k \,
\psi^\dagger_{\alpha k\sigma}(R)\,\psi^\dagga_{\alpha k\sigma}(R)\,,
\ee
where
$\psi^\dagger_{\alpha k\sigma}(R)$ where $\psi^\dagger_{\alpha k\sigma}(R)$ are the scattering waves constructed 
with renormalized hybridization $R\,V_k$, which depend on the retarded impurity Green's function 
\be
g_{dR}(\epsilon)=\frac{1}{\epsilon-\mu+iR^2\Gamma}\,.
\ee  
We observe that, if $\mu$ is taken to be the value that satisfies the constraint (\ref{duech4}) at equilibrium,  
the ground state $|\Psi_0(\Phi)\rangle$ of the non-equilibrium Hamiltonian
\bea
\hat{\mathcal{H}}^0_{R}(\Phi) &=& 
\sum_{\alpha k\sigma}\epsilon_k \,\psi^\dagger_{\alpha k\sigma}(R)\,\psi^\dagga_{\alpha k\sigma}(R)
\nonumber \\
&+&\, \Phi\,
\sum_{\alpha k\sigma}\frac{\alpha}{2} \,\psi^\dagger_{\alpha k\sigma}(R)\,\psi^\dagga_{\alpha k\sigma}(R)
\eea 
is not such that
\be
\Av{\Psi_0(\Phi)}{\dc_\sigma\da_\sigma}=
n\,,
\label{ragionecasino}
\ee
namely it doesn't satisfy anymore (\ref{duech4}).
The procedure described for $\epsilon_d=0$ should then be modified 
without fulfilling this condition. This forces us to renounce to the mechanism
that eliminates the diverging terms of the variance mentioned in 
Sec.~\ref{varoptsubsec}, and that we interpreted as 
measure of merit of the variational ansatz defined in
Eq.~\eqref{varchoicenoneq}.


\section{Conclusions}\label{conclusions}

We have defined a variational principle based on the minimization of a bias-dependent functional
of the variance for studying the steady-state zero-temperature properties of a general quantum-dot 
driven out of equilibrium through the application of a bias.
We have proposed a similar (although approximated) method based on a ``constrained'' minimization of the energy.

The ideas proposed in this paper 
are mainly inspired by the Hershfield's point of view~\cite{Her.} that
the out-of-equilibrium steady state can be regarded as the equilibrium 
one with an Hamiltonian $\hat{\mathcal{H}}(\Phi)$ that includes an 
effective nonequilibrium term proportional to the bias $\Phi \hat{Y}$.
Our main result is that, equivalently, the steady state can be identified 
by the following conditions:
\begin{itemize}
\item the initial state $\rho_0(\Phi)$ (identified by  $\Phi \hat{Y}_0$) defines a \emph{phase} $\mathcal{I}(\Phi)$ of 
  ``macroscopically equivalent'' states that contains also the steady state $\rho(\Phi)$
\item $\rho(\Phi)$ is the ``only'' stationary state of the correlated Hamiltonian $\mathcal{H}$ in $\mathcal{I}(\Phi)$.
\end{itemize}
Such characterization of the nonequilibrium steady state does \emph{not} 
require the explicit knowledge of the Hershfield's operator $\hat{Y}$, 
and, for this reason, we believe that it constitutes an interesting and useful
formulation of the problem.
It opens, in fact, a new possibility to treat nonequilibrium using equilibrium methods.

In order to test our methods, we have considered the simple single orbital  
Anderson impurity model at half-filling, finding a good qualitative accord with the observed 
behavior in quantum dots for the expected regime of validity.
The choice of the variational space was, in fact, based on the the assumption that the effective Hershfield Hamiltonian 
$\hat{\mathcal{H}}(\Phi)$ describes a local Fermi liquid theory~\cite{0953-8984-16-16-R01} in the Nozi\'eres 
sense.~\cite{Nozieres-Journal-of-Low-Temperature-Physics}

The ideas that we have proposed have the big advantage of being very simple, and
we believe that further developments will enable us to deal with more complicated situations
and variational spaces.

\begin{acknowledgments}
I am truly indebted to Prof. Michele Fabrizio for insightful discussions that
allowed me to clarify several important points related to this work.
Further more I thank Prof. Giovanni Morchio, Prof. Bo Hellsing and 
Hugo Strand for constructive discussions and comments on the manuscript.
\end{acknowledgments}

\appendix*
\section{Derivation of Eq.~\eqref{equationvariance2}}

The variance of the Anderson impurity model~\eqref{Hamch4} respect to 
our Gutzwiller variational function~\eqref{GWFch4} is given by
\begin{widetext}
\bea
&&\sigma_{\mathcal{H}}^2\left[\Psi\right]
=\left(
\Av{\Psi_{0}}{\mathcal{P}_d^\dagger\,\hat{T}^2\,\mathcal{P}_d^\dagga}
\!-\!\Av{\Psi_{0}}{\mathcal{P}_d^\dagger\,\hat{T}\,\mathcal{P}_d^\dagga}^2
\right)
\nonumber\\
&&\quad + \left(
\Av{\Psi_{0}}{\mathcal{P}_d^\dagger\,\hat{T}\hat{V}\,\mathcal{P}_d^\dagga}
\!+\!\Av{\Psi_{0}}{\mathcal{P}_d^\dagger\,\hat{V}\hat{T}\,\mathcal{P}_d^\dagga}
-2\,\Av{\Psi_{0}}{\mathcal{P}_d^\dagger\,\hat{T}\,\mathcal{P}_d^\dagga}
\Av{\Psi_{0}}{\mathcal{P}_d^\dagger\,\hat{V}\,\mathcal{P}_d^\dagga}
\right)
\nonumber\\
&&\quad + \left(
\Av{\Psi_{0}}{\mathcal{P}_d^\dagger\,\hat{V}^2\,\mathcal{P}_d^\dagga}
\!-\!\Av{\Psi_{0}}{\mathcal{P}_d^\dagger\,\hat{V}\,\mathcal{P}_d^\dagga}^2
\right)
\nonumber\\
&&\quad 
+ \left(
\Av{\Psi_{0}}{\mathcal{P}_d^\dagger\,\hat{V}\hat{U}\,\mathcal{P}_d^\dagga}
\!+\!\Av{\Psi_{0}}{\mathcal{P}_d^\dagger\,\hat{U}\hat{V}\,\mathcal{P}_d^\dagga}
-2\,\Av{\Psi_{0}}{\mathcal{P}_d^\dagger\,\hat{U}\,\mathcal{P}_d^\dagga}
\Av{\Psi_{0}}{\mathcal{P}_d^\dagger\,\hat{V}\,\mathcal{P}_d^\dagga}
\right)
\nonumber\\
&&\quad + \left(
\Av{\Psi_{0}}{\mathcal{P}_d^\dagger\,\hat{T}\hat{U}\,\mathcal{P}_d^\dagga}
\!+\!\Av{\Psi_{0}}{\mathcal{P}_d^\dagger\,\hat{U}\hat{T}\,\mathcal{P}_d^\dagga}
-2\,\Av{\Psi_{0}}{\mathcal{P}_d^\dagger\,\hat{T}\,\mathcal{P}_d^\dagga}
\Av{\Psi_{0}}{\mathcal{P}_d^\dagger\,\hat{U}\,\mathcal{P}_d^\dagga}
\right)
\nonumber\\
&&\quad + \left(\!
\Av{\Psi_{0}}{\mathcal{P}_d^\dagger\,\hat{U}^2\,\mathcal{P}_d^\dagga}
\!-\!\Av{\Psi_{0}}{\mathcal{P}_d^\dagger\,\hat{U}\,\mathcal{P}_d^\dagga}\!^2
\right)
\label{va1}
\eea
\end{widetext}

Our calculation can be considerably simplified by the
following considerations.
\begin{itemize} 
\item The variance of the renormalized Hamiltonian
\bea
\mathcal{H}_0^R &=& \hat{T}+\hat{V}_R
\label{hrinfree}
\eea
on the function $|\Psi_0\rangle$ 
\bea
\sigma_{\mathcal{H}_0^R}\left[\Psi_{0}\right]
\!\!&=&\!\!\left(
\Av{\Psi_{0}}{\hat{T}^2}
-\Av{\Psi_{0}}{\hat{T}}^2
\right)
\nonumber\\
\!\!&+&\!\!R\left(
\Av{\Psi_{0}}{\hat{T}\hat{V}}
+\Av{\Psi_{0}}{\hat{V}\hat{T}}
\right.
\nonumber\\
&&-\left.
2\,\Av{\Psi_{0}}{\hat{T}}
\Av{\Psi_{0}}{\hat{V}}
\right)
\nonumber\\
\!\!&+&\!\!R^2\left(
\Av{\Psi_{0}}{\hat{V}^2}
-\Av{\Psi_{0}}{\hat{V}}^2
\right)
\label{va2}
\eea
is zero, because $|\Psi_{0}\rangle$ is, by definition, 
the ground state of $\mathcal{H}_0^R$.

\item Our variational function satisfies the 
Gutzwiller constraint defined in Eq.~\eqref{duech4}, so that the following 
equations holds:
\bea
\Av{\Psi_{0}}{\mathcal{P}_d^\dagger\,\hat{T}\,\mathcal{P}_d^\dagga}&=&
\Av{\Psi_{0}}{\hat{T}}\nonumber\\
\Av{\Psi_{0}}{\mathcal{P}_d^\dagger\,\hat{V}\,\mathcal{P}_d^\dagga}&=&
R\,\Av{\Psi_{0}}{\hat{V}}
\eea

\item A direct calculation shows that 
\bea
\Av{\Psi_{0}}{\mathcal{P}_d^\dagger\,\hat{T}\hat{V}\,\mathcal{P}_d^\dagga}
\!\!&=&\!\!R\,\Av{\Psi_{0}}{\hat{T}\hat{V}}
\nonumber\\
\Av{\Psi_{0}}{\mathcal{P}_d^\dagger\,\hat{V}\hat{T}\,\mathcal{P}_d^\dagga}
\!\!&=&\!\!R\,\Av{\Psi_{0}}{\hat{V}\hat{T}}
\eea
\bea
&&\Av{\Psi_{0}}{\mathcal{P}_d^\dagger\,\hat{T}\hat{U}\,\mathcal{P}_d^\dagga}
+\Av{\Psi_{0}}{\mathcal{P}_d^\dagger\,\hat{U}\hat{T}\,\mathcal{P}_d^\dagga}
\nonumber\\
&&\quad\!\!\!\!\!
-2\,\Av{\Psi_{0}}{\mathcal{P}_d^\dagger\,\hat{T}\,\mathcal{P}_d^\dagga}
\Av{\Psi_{0}}{\mathcal{P}_d^\dagger\,\hat{U}\,\mathcal{P}_d^\dagga}\!=\!0
\eea
\bea
&&\Av{\Psi_{0}}{\mathcal{P}_d^\dagger\,\hat{V}\hat{U}\,\mathcal{P}_d^\dagga}
+\Av{\Psi_{0}}{\mathcal{P}_d^\dagger\,\hat{U}\hat{V}\,\mathcal{P}_d^\dagga}
\label{aterm}
\nonumber\\
&&\quad-2\,\Av{\Psi_{0}}{\mathcal{P}_d^\dagger\,\hat{U}\,\mathcal{P}_d^\dagga}
\Av{\Psi_{0}}{\mathcal{P}_d^\dagger\,\hat{V}\,\mathcal{P}_d^\dagga}
\nonumber\\
&&=\frac{U}{2}\,\lambda_{1}\lambda_{0,2}\left(1-\lambda_{0,2}^2\right)
\Av{\Psi_{0}}{\hat{V}}
\eea
\bea
&&\Av{\Psi_{0}}{\mathcal{P}_d^\dagger\,\hat{U}^2\,\mathcal{P}_d^\dagga}
-\Av{\Psi_{0}}{\mathcal{P}_d^\dagger\,\hat{U}\,\mathcal{P}_d^\dagga}^2
\nonumber\\
&&=\left(\frac{U}{2}\right)^2\,
\frac{\lambda_{0,2}^2}{2}\left(1-\frac{\lambda_{0,2}^2}{2}\right)
\eea

\end{itemize}

Taking the difference between Eq.~\eqref{va1} and Eq.~\eqref{va2} 
and using the above equations we obtain that
\bea
\sigma^2_{\mathcal{H}}\left[\Psi\right]
&=&\left(
\Av{\Psi_{0}}{\mathcal{P}_d^\dagger\mathcal{P}_d^\dagga\,\hat{T}^2}
-\Av{\Psi_{0}}{\hat{T}^2}
\right)
\nonumber\\
&+&\left(
\Av{\Psi_{0}}{\mathcal{P}_d^\dagger\,\hat{V}^2\,\mathcal{P}_d^\dagga}
-
R^2\,
\Av{\Psi_{0}}{\hat{V}^2}
\right)
\nonumber\\
&+&\frac{U}{2}\,\lambda_{1}\lambda_{0,2}\left(1-\lambda_{0,2}^2\right)
\Av{\Psi_{0}}{\hat{V}}
\nonumber\\
&+&\left(\frac{U}{2}\right)^2\,
\frac{\lambda_{0,2}^2}{2}\left(1-\frac{\lambda_{0,2}^2}{2}\right)\,.
\label{va}
\eea

Let us consider now the first term in Eq.~\eqref{va}, which
is equal to the sum of all the Wick contractions in which
the operators $\mathcal{P}^\dagger\mathcal{P}$ and $\hat{T}^2$
are connected by two or four ``legs''.

A direct calculation shows that
\bea
\mathcal{P}_d^\dagger\mathcal{P}_d^\dagga&=&\lambda_{0,2}^2
-\left(\lambda_{0,2}^2-\lambda_{1}^2\right)
\left(\dc_\uparrow\da_\uparrow+\dc_\downarrow\da_\downarrow\right)\nonumber\\
&+& 2\left(\lambda_{0,2}^2-\lambda_{1}^2\right)
\dc_\uparrow\da_\uparrow\,\dc_\downarrow\da_\downarrow\,.
\label{pp}
\eea
Using Eq.~\eqref{pp} it can be easily verified that the 
sum of all the two-legs contraction between 
$\mathcal{P}_d^\dagger\mathcal{P}_d$ and $\hat{T}^2$ is zero.
The four legs contribution can be calculated using Wick's theorem. 
The result is 
\bea
&&\Av{\Psi_0}{\mathcal{P}_d^\dagger\mathcal{P}_d^\dagga\,\hat{T}^2}
-\Av{\Psi_0}{\hat{T}^2}\nonumber\\
&&\quad\quad=\left(\lambda_{0,2}^2-\lambda_{1}^2\right)\mathcal{A}^2(\mathcal{P}_d)\,,
\eea
where
\be
\mathcal{A}(\mathcal{P}_d)\equiv\sum_{k\alpha\sigma}\epsilon_k
\left|
\Av{\Psi_0}{\dc_\sigma\ca_{\alpha k\sigma}}
\right|^2\,.
\ee

Let us consider the second term in Eq.~\eqref{va}. 
It can be verified that
\bea
&&\Av{\Psi_{0}}{
\mathcal{P}_d^\dagger\,\hat{V}^2\,\mathcal{P}_d^\dagga}
-R^2\,\Av{\Psi_{0}}{
\mathcal{P}_d^\dagger\,\hat{V}^2\,\mathcal{P}_d^\dagga} \nonumber\\
&\dagga&\quad=
\left(1-R^2\right)
\left(\sum_{k\alpha}\frac{V_k^2}{\Omega}+3\,\frac{\mathcal{V}^2(\mathcal{P}_d)}{R^2}\right)
\,;
\eea
where 
\be
\mathcal{V}(\mathcal{P}_d)
\equiv \sum_{k\alpha\sigma}\frac{R\,V_k}{\sqrt{\Omega}}\,
\Av{\Psi_{0}}{\cc_{\alpha k\sigma}\da_\sigma}+c.c.\;.
\label{defV}
\ee

The above calculations lead to the following expression for the variance:
\bea
\sigma^2_{\mathcal{H}}\left[\Psi\right]
&=&
\left(\lambda_{0,2}^2-\lambda_{1}^2\right)\mathcal{A}^2(\mathcal{P}_d)\nonumber\\
&+&\left(1-R^2\right)
\left(\sum_{k\alpha}\frac{V_k^2}{\Omega}+3\,\frac{\mathcal{V}^2(\mathcal{P}_d)}{R^2}\right)\nonumber\\
&+&\frac{U}{2}\left(1-\lambda_{0,2}^2\right)
\mathcal{V}(\mathcal{P}_d)\nonumber\\
&+&\left(\frac{U}{2}\right)^2\,
\frac{\lambda_{0,2}^2}{2}\left(1-\frac{\lambda_{0,2}^2}{2}\right)
\nonumber\\
&\equiv&
\sigma_{\mathcal{H}_0}^2\left[\Psi\right]
+\delta\sigma_{int}^2\left[\Psi\right]
\,.
\label{equationvariance}
\eea

Notice that the functional relation~\eqref{relRhrin} between 
$|\Psi_{0}\rangle$ and $\mathcal{P}_d$ has been responsible of
the cancellation of the $\hat{T}\hat{V}$ terms in Eq.~\eqref{va1},
that are extensive quantities.

It can be easily proven that the minimum condition of $\sigma_{\mathcal{H}_0}^2$ in the not projected 
state implies that 
\be
\mathcal{A}^2(\mathcal{P}_d)=0\,.
\ee
The explicit value of $\mathcal{V}(\mathcal{P}_d)$ can be simply obtained replacing 
$\Gamma$ with $z\Gamma$ (see Eq.~\eqref{gutzqprelch4}) in Eq.~\eqref{hybgain}. 

If, for simplicity, we assume that
\bea
&&V_k=V_{k'}\quad\forall k,k'\nonumber\\
&&\Gamma\ll W=1\,,
\eea
it can be easily verified that the variance is given by 
\bea
\sigma^2_{\mathcal{H}}\left[\Psi\right]
&=&\left(1-z\right)
\left(\frac{\Gamma}{2}+\frac{12}{\pi^2}\,\Gamma^2 z \log^2\left(z^2\Gamma^2\right)\right)\nonumber\\
&+&\frac{U}{\pi}\, z \sqrt{1-z}\,
\Gamma \log\left(z^2\Gamma^2\right)+\frac{z U^2}{16}
\,,
\eea
that coincides with Eq.~\eqref{equationvariance2}.

\bibliographystyle{apsrev}

\end{document}